\documentclass[secnumarabic,twocolumn,graphics,floatfix,tightenlines,aps,prx,superscriptaddress,longbibliography]{revtex4-2}

\pdfoutput=1

\usepackage[utf8]{inputenc}
\usepackage[english]{babel}
\usepackage[T1]{fontenc}
\usepackage{amsmath}

\usepackage{graphicx}
\usepackage{dcolumn}
\usepackage{booktabs}

\usepackage{soul}
\sethlcolor{pink}

\usepackage[dvipsnames]{xcolor}
\usepackage{tikz}
\usetikzlibrary{arrows}
\usetikzlibrary{calc}
\usepackage[caption=false]{subfig}
\graphicspath{ {./images/} }
\usepackage{physics}
\usetikzlibrary{shapes.geometric}
\usetikzlibrary{positioning}

\usepackage{url}
\usepackage{hyperref}
\hypersetup{
    colorlinks=true,
    linkcolor=blue,
    citecolor=blue, 
    filecolor=magenta,      
    urlcolor=blue,
}

\usepackage{mleftright} 

\renewcommand{\eqref}[1]{\mbox{Eq.~(\ref{#1})}}

\newcommand{\figpanel}[2]{Fig.~\hyperref[#1]{\ref*{#1}(#2)}}
\newcommand{\figpanels}[3]{Fig.~\hyperref[#1]{\ref*{#1}(#2)-(#3)}}
\newcommand{\figpanelNoPrefix}[2]{\hyperref[#1]{\ref*{#1}(#2)}}

\usepackage{units}

\begin{document}

\title{Data-driven decoding of quantum error correcting codes using graph neural networks}

\author{Moritz Lange}
\affiliation{Department of Physics, University of Gothenburg, Gothenburg, Sweden}

\author{Pontus Havstr\"om}
\affiliation{Department of Physics, University of Gothenburg, Gothenburg, Sweden}

\author{Basudha Srivastava}
\affiliation{Department of Physics, University of Gothenburg, Gothenburg, Sweden}

\author{Isak Bengtsson}
\affiliation{Department of Physics, Chalmers University of Technology,  Gothenburg, Sweden}

\author{Valdemar Bergentall}
\affiliation{Department of Physics, University of Gothenburg,  Gothenburg, Sweden}

\author{Karl Hammar}
\affiliation{Department of Physics, University of Gothenburg, Gothenburg, Sweden}

\author{Olivia Heuts}
\affiliation{Department of Physics, University of Gothenburg, Gothenburg, Sweden}

\author{Evert van Nieuwenburg}
\email[]{evert.vn@lorentz.leidenuniv.nl}
\affiliation{Leiden Inst. of Advanced Computer Science, Leiden University, Leiden, Netherlands}

\author{Mats Granath}
\email[]{mats.granath@physics.gu.se}
\affiliation{Department of Physics, University of Gothenburg, Gothenburg, Sweden}


\captionsetup[subfigure]{labelformat=empty}

\begin{abstract}

To leverage the full potential of quantum error-correcting stabilizer codes it is crucial to have an efficient and accurate decoder. 
Accurate, maximum likelihood, decoders are computationally very expensive whereas decoders based on more efficient algorithms give sub-optimal performance. In addition, the accuracy will depend on the quality of models and estimates of error rates for idling qubits, gates, measurements, and resets, and will typically assume symmetric error channels. In this work,  we explore a model-free, data-driven, approach to decoding, using a graph neural network (GNN). The decoding problem is formulated as a graph classification task in which a set of stabilizer measurements is mapped to an annotated detector graph for which the neural network predicts the most likely logical error class. We show that the GNN-based decoder can outperform a matching decoder for circuit level noise on the surface code given only the simulated data, while the matching decoder is given full information of the underlying error model. Although training is computationally demanding, inference is fast and scales approximately linearly with the space-time  volume of the code. We also find that we can use large, but more limited, datasets of real experimental data for the repetition code, giving decoding accuracies that are on par with minimum weight perfect matching. The results show that a purely data-driven approach to decoding may be a viable future option for practical quantum error correction, which is competitive in terms of speed, accuracy, and versatility.

\end{abstract}

\maketitle


\section{Introduction}

Quantum Error Correction (QEC) is foreseen to be a vital component in the development of practical quantum computing~\cite{Shor1995SchemeMemory, Steane1996ErrorTheory, Gottesman1997StabilizerCorrection, Terhal2015QuantumMemories, girvin2021introduction}. The need for QEC arises due to the susceptibility of quantum information to noise, which can rapidly accumulate and corrupt the final output. Unlike noise mitigation schemes where errors are reduced by classical post-processing \cite{kim2023evidence,temme2017error,li2017efficient}, QEC methods encode quantum information in a way that allows for the detection and correction of errors without destroying the information itself. A prominent framework for this is topological stabilizer codes, such as the surface code, for which the logical failure rates can be systematically suppressed by increasing the size of the code if the intrinsic error rates are below some threshold value~\cite{Bravyi1998QuantumBoundary, Dennis2002TopologicalMemory, Kitaev2003Fault-tolerantAnyons, Raussendorf2007Fault-TolerantDimensions, Fowler2012SurfaceComputation}. 

Stabilizer codes are based on a set of commutative, typically local, measurements that project an $n$-qubit state to a lower dimensional code space representing one or more logical qubits. 
Errors take the state out of the code space and are then indicated by a syndrome, corresponding to stabilizer violations. 
The syndrome needs to be interpreted in order to gauge whether a logical bit or phase flip may have been incurred on the logical qubit. 
Interpreting the syndrome, to predict the most likely logical error, requires both a decoder algorithm and, traditionally, a model of the qubit error channels. 
The fact that measurements may themselves be noisy, makes this interpretation additionally challenging \cite{Dennis2002TopologicalMemory,Fowler2012SurfaceComputation}. 

Efforts are under way to realize stabilizer codes experimentally using various qubit architectures~\cite{Kelly2015StateCircuit,Takita2017ExperimentalQubits,PhysRevA.97.052313, wootton2020benchmarking,Andersen2020RepeatedCode, Satzinger2021Realizing, Egan2021Fault-tolerantQubit, Chen2021ExponentialCorrection, Erhard2021EntanglingSurgery, ryananderson2021realization, marques2021logicalqubit, Postler_2022, krinner2022realizing,Bluvstein2021,google2023suppressing,moses2023race,sundaresan2023demonstrating}.  
In \cite{google2023suppressing}, code distance 3 and 5 surface codes were implemented, using 17 and 49 superconducting qubits, respectively. 
After initialization of the qubits, repeated stabilizer measurements are performed over a given number of cycles capped by a final round of single qubit measurements. 
The results are then compared with the initial state to determine whether errors have caused a logical bit- (or phase-) error. 
The decoder analyses the collected sets of syndrome measurements in post-processing, where the fraction of correct predictions gives a measure of the logical accuracy. 
The better the decoder, the higher the coherence time of the logical qubit, and in \cite{google2023suppressing} a computationally costly tensor network based decoder was used to maximize the logical fidelity of the distance 5 code compared to the distance 3 code. However, with the objective of moving from running and benchmarking a quantum memory to using it for universal quantum computation, it will be necessary to do error correction both with high accuracy and in real time.

\begin{figure}
\centering
\includegraphics[width=\linewidth]{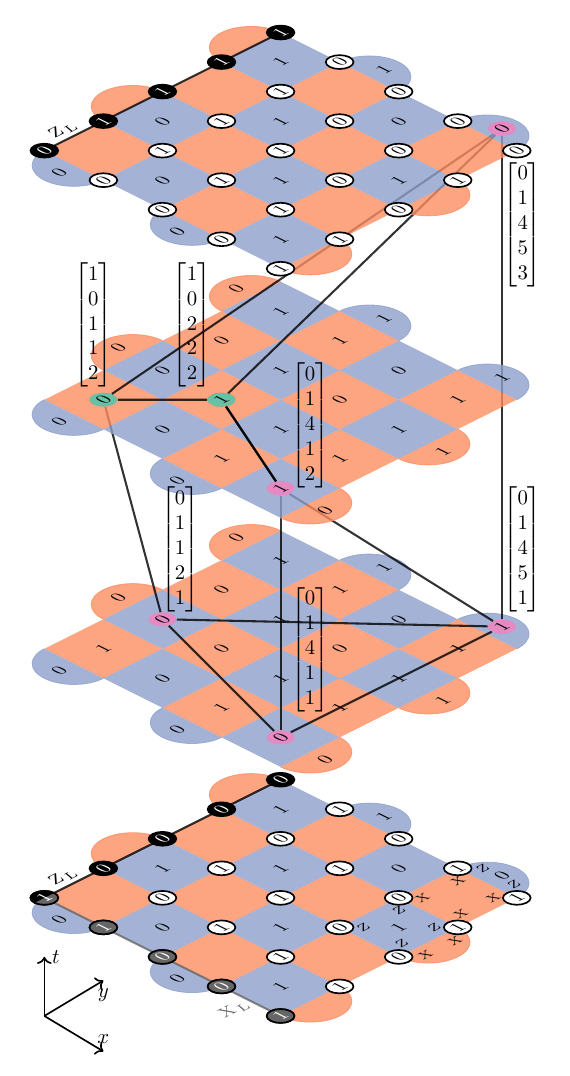}
\caption{Memory experiment on the distance $d=5$ surface code. Data qubit initialization is followed by $d_t=2$ stabilizer measurement rounds and a final data qubit measurement round. Data qubits are on the vertices of plaquettes (circles, shown in the bottom and top planes). Ancilla qubits (not shown) at the center of plaquettes provide stabilizer measurements outcomes. The detector graph has nodes corresponding to changes in stabilizers from the previous time step. (Not all edges shown.) Nodes are annotated by the type of stabilizer and the space-time coordinate. The label, here $\lambda_Z=1$, corresponding to a change of $\langle Z_L\rangle$, measured along the northwest edge. Also shown, bottom layer, are some example stabilizers, and the logical $X_L$ (not measured).}
\label{fig:memory_experiment}
\end{figure}

\begin{figure*}
\centering
\includegraphics[width=\linewidth]{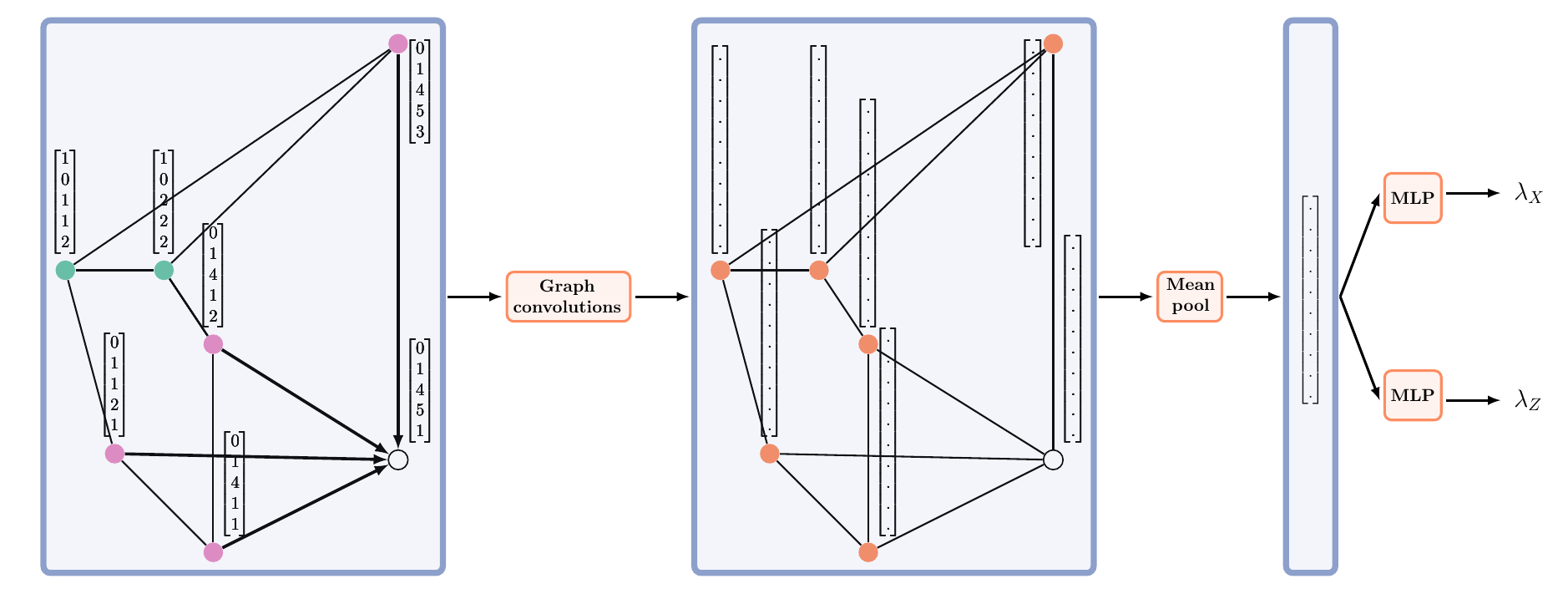}
\caption{Schematic of the GNN decoder. It takes as input an annotated detector graph, c.f. Figure \ref{fig:memory_experiment}. Several layers of graph convolutional operations (following Eqn.~\ref{Eq:graphconv}) transform each node feature vector. (The empty circle shows the message passing to this particular node from  neighboring nodes on the graph.) Next, a mean-pooling operation averages all the node feature vectors into a single graph embedding, which is independent of the size of the graph. Finally, the latter is passed through two separate dense networks to give two binary class predictors, corresponding to the logical $X$ and $Z$ labels, respectively. (For details see Appendix \ref{sec:App_A}.)    
}
\label{fig:GNN_scheme}
\end{figure*}

In the present work, we explore the viability of using a purely data-driven approach to decoding, based on the potential of generating large amounts of experimental data. 
We use a graph neural network (GNN) which is well suited for addressing this type of data. 
Namely, a single data point, as in \cite{google2023suppressing}, consists of a set of  ``detectors'', i.e., changes in stabilizer measurements from one cycle to the next, together with a label indicating the measured logical bit- or phase-flip error. 
This can be represented as a labeled graph with nodes that are annotated by the information on the type of stabilizer and the space-time position of the detector, as shown in Figure \ref{fig:memory_experiment}. 
The maximum degree of the graph can be capped based on removing edges between distant detectors, keeping only a fixed maximum number of neighboring nodes. 
The latter ensures that each network layer in the GNN (see Figure \ref{fig:GNN_scheme}) performs a number of matrix multiplications that scales linearly with the number of nodes, i.e., linearly with the number of stabilizer measurements and the overall error rate. 
We have trained this decoder on simulated data for the surface code using Stim~\cite{gidney2021stim} as well as real experimental data on the repetition code~\cite{google2023suppressing}. 
For both of these, the decoder is on par with, or outperforms, state-of-the-art matching decoders~\cite{higgott2021pymatching,Higgott2022beliefmatching}, suggesting that with sufficient data and a suitable neural network architecture, model-free machine learning based decoders trained on experimental data can be competitive for future implementations of quantum error-correcting stabilizer codes. 

\section{Stabilizer codes and decoding}
\label{sec:stab}

A stabilizer code is defined through a set of commuting operators constructed from products of Pauli operators acting on a Hilbert space of $n$ data qubits \cite{Gottesman1997StabilizerCorrection}.
With $n_S$ independent stabilizers the Hilbert space is split into sectors of dimension $2^{n-n_S}$, specified by the parity under each stabilizer. 
For concreteness we will consider the case $n_S=n-1$, such that each of the sectors represent a single qubit degree of freedom. Each syndrome measurement is performed with the help of an ancilla qubit following a small entangling circuit with the relevant data qubits. The measured state of the ancilla qubits provide a syndrome 
$S=\{s_i, i=1,...,n_S\, | \in 0,1 \}$, and projects the density matrix of the $n$ qubit state into a single 2-dimensional block, a Pauli frame~\cite{knill2005quantum,deBeaudrap2020zxcalculusis}. Given uncertainties in the measurements, a number of rounds are typically performed before the information is interpreted by means of a decoder. 

Defining a pair of anticommuting operators $Z_L$ and $X_L$ that commute with the stabilizer group, provides the logical computational space through $Z_L|0\rangle_L=|0\rangle_L$ and $|1\rangle_L=X_L|0\rangle_L$. Assuming a fixed pair of logical operators for a given code defines the corresponding logical states in each Pauli frame. Thus, a number of subsequent rounds of stabilizer measurements, during which the code is affected by decoherence, transforms the density matrix from the initial state 
\begin{equation}
   \rho=\sum_{i,j\in \{0,1\}}\rho_{ij}|i\rangle_L\langle j|_L 
\end{equation}
to the final state 
\begin{equation}
   \rho'=\sum_{i,j\in \{0,1\}}\rho'_{ij}|i\rangle'_L\langle j|'_L\,,
\end{equation}
where $|0/1\rangle_L$ ($|0/1\rangle'_L$) are the logical qubit states in the initial (final) Pauli frame. The logical error channel is approximated by   
\begin{eqnarray}
   &\rho&\rightarrow \rho' = \epsilon_L(\rho) \label{Eq:rho}\\
    &=& (1-P)\tilde{\rho}+P_X X_L\tilde{\rho} X_L+P_Z Z_L\tilde{\rho} Z_L+P_Y Y_L\tilde{\rho} Y_L\,,\nonumber
\end{eqnarray}
with $Y_L=-iZ_LX_L$ and $P=\sum_{i=X,Y,Z}P_i$. Here $\tilde{\rho}=C(s,s')\rho C(s,s')$, where $C(s,s')|0/1\rangle_L=|0/1\rangle'_L$, is an arbitrary Pauli string that effectuates the change of Pauli frame and commutes with the logical operators.  
In general there may be additional non-symmetric channels (see for example \cite{Satzinger2021Realizing}), but we will assume that the data (as in \cite{google2023suppressing}) does not resolve such channels. 


The probabilities of logical error, $P_i$, will be quantified by the complete set of syndrome measurements and depend on single and multi-qubit error channels as well as measurement and reset errors. It is the task of the decoder to quantify these in order to maximize the effectiveness of the error correction. 
Traditionally this is done through computational algorithms that use a specific error model. 
The framework that most decoders are based on uses independent and identically distributed symmetric noise acting on individual qubits, possibly, for circuit-level noise, complemented by two-qubit gate errors, faulty measurements and ancilla qubit reset errors.  
Maximum-likelihood decoders~\cite{wootton2012,Hutter2014EfficientCode,Bravyi2014EfficientCode,Hammar_2022,pryadko2020maximum, Chubb2021TensorNetwork} aim to explicitly account for all possible error configurations that are consistent with the measured syndromes, with their respective probabilities given by the assumed error model. 
The full set of error configurations fall in four different cosets that map to each other by the logical operators of the code, thus directly providing an estimate of the probabilities $P_i$ that is limited only by the approximations involved in the calculation and the error model. Even though such decoders may be useful for benchmarking and optimizing the theoretical performance of stabilizer codes~\cite{google2023suppressing}, they are computationally too demanding for real time operation, even for small codes.        

\begin{figure}
\centering
\includegraphics[width=\linewidth]{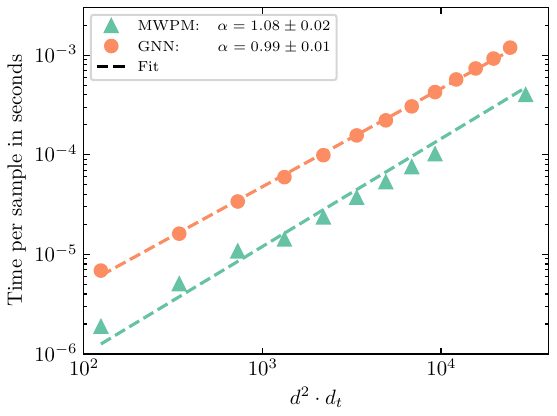}
\caption{Scaling of average decoding time per syndrome versus code volume $d^2d_t$ for GNN and MWPM using PyMatching 2~\cite{pymatchingv2}. Circuit level noise sampled at $p = 10^{-3}$. Dotted lines show a regression according to the ansatz: $T = C \cdot (d^2\cdot d_t)^\alpha$, with the GNN showing sub-linear scaling.}
\label{fig:scaling}
\end{figure}

An alternative type of decoder is based on the minimum weight perfect matching (MWPM) algorithm~\cite{edmonds1965paths, dennis2002topological, wang2009threshold, PhysRevA.83.020302, fowler2015minimum, Brown2022conservationlaws}. 
The objective is to find the single, most likely, configuration of errors consistent with the set of measured stabilizers. 
Detectors are mapped to nodes of a graph with edges that are weighted by the (negative log) probability of the pair of nodes. 
For codes where nodes appear in pairs (such as the repetition or surface code), the most likely error corresponds to pairwise matching such that the total weight of the edges is minimized. 
This algorithm is fast, in practice scaling approximately linearly with the size of the graph (see Figure \ref{fig:scaling}).  
Nevertheless, it has several short-comings that limits accuracy and applicability: 1) Approximate handling of crossing edges (such as coinciding X and Z errors) means that the effective error model is oversimplified. 2) Degeneracies of less likely error configurations are ignored. 3) For models where a single error may give rise to more than two detector events, more sophisticated algorithms are needed~\cite{Delfosse2014ColorCode, Stephens2014, delfosse2017almost, Tuckett2020FaultTolerant, Sahay2022, Berent2023maxsat, Benhemou2023}. 
These shortcomings can be partially addressed by more sophisticated approaches such as counting multiplicity or using belief propagation~\cite{delfosse2014decoding, criger2018multi, Higgott2022beliefmatching, Caune2023BP}, but often at the cost of added computational complexity. Other examples of decoder algorithms are based on decoding from small to large scale, such as cellular-automata \cite{herold2015cellular,kubica2018cellular,Miguel2023}, renormalization group \cite{duclos2010fast}, or union-find \cite{delfosse2017almost,PhysRevA.102.012419}. The latter, in particular, is very efficient, but at the cost of sub-optimal performance. 

\subsection{Related work}
A number of different deep learning based decoder algorithms have also been formulated, based on supervised learning, reinforcement learning, and genetic neural algorithms~\cite{torlai2017neural, krastanov2017deep, varsamopoulos2017decoding, baireuther2018machine, breuckmann2018scalable,Baireuther_2019, chamberland2018deep,nautrup2018optimizing,maskara2019advantages, Ni2020neuralnetwork, sweke2020reinforcement,andreasson2018quantum, colomer2020reinforcement, Fitzek_DRL,gicev2021scalable,bhoumik2021efficient,10.21468/SciPostPhys.11.1.005,Meinerz_2022,Overwater_2022,chamberland2022techniques,zhang2023scalable}. Focusing on the works on the surface code and based on supervised learning, these can roughly be separated according to whether they primarily consider perfect stabilizers~\cite{torlai2017neural,krastanov2017deep,varsamopoulos2017decoding,Ni2020neuralnetwork,gicev2021scalable,bhoumik2021efficient,Overwater_2022}, or include measurement noise or circuit-level noise~\cite{baireuther2018machine,chamberland2018deep,Meinerz_2022,chamberland2022techniques,zhang2023scalable}, and whether they are purely data-driven~\cite{torlai2017neural,varsamopoulos2017decoding,baireuther2018machine,chamberland2018deep,Overwater_2022,zhang2023scalable} or involve some auxiliary, model-informed, algorithm or multi-step reduction of decoding~\cite{krastanov2017deep,chamberland2022techniques,Ni2020neuralnetwork,gicev2021scalable,bhoumik2021efficient,Meinerz_2022}. 

The present work is in the category, realistic (circuit-level) noise, and purely data-driven. It is distinguished primarily in that we 1) Use graph neural networks and graph structured data, and 2) Train and test the neural network decoder on real experimental data. In addition, as in several of the earlier works~\cite{chamberland2018deep,baireuther2018machine,Baireuther_2019}, we emphasize the use of a model-free, purely data-driven, approach.  By using experimental stabilizer data, the approximations of traditional model-based decoder algorithms can be avoided. The fact that the real error channels at the qubit level may be asymmetric, due to amplitude damping, have long-range correlations, or involve leakage outside the computational space, is intrinsic to the data. 
This is also in contrast to other data-driven approaches~\cite{wagner2022pauli,PhysRevLett.128.110504,PhysRevLett.130.200601,google2023suppressing,Chen2021ExponentialCorrection} that use stabilizer data to learn the detailed Pauli channels, optimize a decoder algorithm through the edge weights of a matching decoder, or the individual qubit and measure error rates of a tensor network based decoder, as these are all constrained by a specific error model.

\subsection{Repetition code and surface code}

The decoder formalism that we present in this work can be applied to any stabilizer code, requiring only a dataset of measured (or simulated) stabilizers, together with the logical outcomes. 
Nevertheless, to keep to the core issues of training and performance we consider only two standard scalable stabilizer codes: the repetition code and the surface code.

The bit-flip detecting repetition code is defined on a one-dimensional grid of qubits with neighboring pair-wise $Z_i\otimes Z_{i+1}$ stabilizers. 
In the Pauli frame with all $+1$ stabilizers, the code words are $|0\rangle_L=|0\rangle^{\otimes n}$ and $|1\rangle_L=|1\rangle^{\otimes n}$. 
Consider a logical qubit state $|\psi\rangle=\alpha |0\rangle_L + \beta |1\rangle_L$, with complex amplitudes $|\alpha|^2+|\beta|^2=1$. 
The logical bit-flip operator is given by $X_L=\bigotimes_{i}X_i$, which sets the code distance $d_X=n$. 
Assuming perfect stabilizer measurements and independent and identically distributed single qubit bit-flip error probabilities, decoding the repetition code is trivial. 
For any set of stabilizer violations, i.e., odd parity outcomes, there are only two consistent configurations of errors that map to each other by acting with $X_L$. 
A decoder (maximum-likelihood in the case of this simple error model) would suggest the one with fewer errors. 
The repetition code, set up to detect bit-flip errors, is insensitive to phase flip errors, as is clear from the fact that a  phase-flip ($Z$) error on a single qubit also gives a phase-flip error ($\beta\rightarrow -\beta$) on the logical qubit, corresponding to a code distance $d_Z=1$. 
To detect and correct both bit- and phase-flip errors we need a more potent code, the most promising of which may be the surface code.

We consider the qubit-efficient ``rotated'' surface code \cite{Bombin2007OptimalStudy, Tomita2014, Tuckett2019Tailored} (see Figure \ref{fig:memory_experiment}), constructed from weight-4, $Z^{\otimes 4}$ and $X^{\otimes 4}$, stabilizers (formally stabilizer generators), with complementary weight-2 stabilizers on the boundary. 
On a square grid of $d\times d$ data qubits, the $d^2-1$ stabilizers give one logical qubit. 
We define the logical  
operator $X_L$ as a string of $X$'s on the southwest edge, and a string of $Z$'s on the northwest edge, as shown in  Figure \ref{fig:memory_experiment}.
These are the two (unique up to products of stabilizers) lowest weight operators that commute with the stabilizer group, without being part of said group. 

Stabilizer measurements are performed by means of  entangling circuits between the data qubits and an ancilla qubit. Assuming hardware with one ancilla qubit per stabilizer, and the appropriate gate schedule, these can all be measured simultaneously, corresponding to one round of stabilizer measurements. 

\subsection{Memory experiments on the surface code}
\label{sec:memory_exp}
To train and test our decoder we consider a real or simulated experimental setup, illustrated schematically in Figure \ref{fig:memory_experiment}, to benchmark a surface code as a quantum memory. The following procedure can be used for any stabilizer code:

\begin{itemize}
\item Initialize the individual qubits: Data qubits in a fixed or random configuration in the computational basis $|0\rangle$ and $|1\rangle$. Ancilla qubits in $|0\rangle$. The initial data qubit configuration is viewed as a 0'th round of measurements that initialize the $Z$-stabilizers $s_{Z,i,t=0}$.
This also corresponds to an effective measurement $\langle Z_L\rangle_{t=0}=\prod_{i\in Z_L}Z_i=\pm 1$. (Northwest row of qubits in Figure \ref{fig:memory_experiment}.) 
\item A first round, $t=1$, of actual stabilizer measurements is performed, with outcomes $s_{Z,i,t=1}$ and $s_{X,i,t=1}$. This provides the first round of Z-detectors corresponding to changes in $s_{Z,i}$ from the inititalization step. The X-stabilizers $s_{X,i,t=1}$ have randomized outcome, projecting to an even or odd parity state over the four (or two) qubits in the Hadamard ($|+\rangle$, $|-\rangle$) basis. The value of these stabilizers form the reference for subsequent error detecting measurements of the X-stabilizers. Ancilla qubits are reset to 0 after this and subsequent rounds. 
\item Subsequent rounds $t=2,...,d_t$ of Z and X stabilizer measurements provide the input for corresponding detectors based on changes from the previous round. 

\item Finally, data qubits are measured individually in the Z-basis, 
which provides a final measurement, $\langle Z_L\rangle_{t=d_t+1}$.
The measurements also provide Z-stabilizers, which, being calculated from the actual qubit outcomes rather than by measuring an ancilla, are perfect stabilizers by definition. 
\end{itemize}

The outlined experiment provides a single data point $D=(\{V_{Z}\},\{V_{X}\},\lambda_Z)$ consisting of a set of Z- and X-detectors $\{V_{Z}\}$ and $\{V_{X}\}$, together with a logical label $\lambda_Z$. The detectors are defined as the non-zero outcomes of
\begin{eqnarray}
    V_{Z,i,t}&=&s_{Z,i,t-1}\oplus s_{Z,i,t}\\
    V_{X,i,t}&=&s_{X,i,t-1}\oplus s_{X,i,t}\,,
\end{eqnarray}
i.e.\ corresponding to a change in a stabilizer measurement from one-time step to the next. In addition to the stabilizer type, each detector is tagged with its space-time coordinate, $(x_i,y_i,t)$, with $0\leq x,y\leq d$ and $1\leq t\leq d_t\pm 1$ for $Z$ and $X$ detectors respectively. 
The logical label is given by
\begin{equation}
    \lambda_Z=\frac{1}{2}|\langle Z_L\rangle_{t=0}-\langle Z_L\rangle_{t=d_t+1}|\in \{0,1\}\,.
\end{equation} 
The probability of $\lambda_Z=1$ is, according to Eqn.~\ref{Eq:rho}, given by $P_X+P_Y$, and the probability of $\lambda_Z=0$ by $P_I+P_Z$, corresponding to a logical bit-flip or not.  

What has been described is a ``memory-Z'' experiment~\cite{gidney2021stim}, i.e., one in which we detect logical bit-flips. 
Qubits are initialized in the computational basis $|0\rangle$ and $|1\rangle$. 
A ``memory-X'' experiment prepares the qubits in the Hadamard basis, with the role of X- and Z-stabilizers reversed. 
Physically, in the lab, one cannot do both experiments in the same run, as $Z_L$ and $X_L$ do not commute. 
This also implies that each data point only has one of the two binary labels, $\lambda_Z$ or $\lambda_X$, even though there is information in the detectors about both labels. 
The neural network will be constructed to predict both labels for a given set of detectors, which implies that the learning framework is effectively that of semi-supervised learning, with partially labeled data. 
Thus, in contrast to a matching based decoder, which breaks the surface code detectors into two independent sets with a corresponding graph for each, the GNN decoder can make use of the complete information. 
This, in addition to the fact that it is not constrained by the limitations of the matching algorithm itself, provides a possible advantage in terms of prediction accuracy.  


We have also assumed that there is no post-processing to remove leakage. Assuming there is some mechanism of relaxation back to the computational qubit subspace, including the last round of measurements, leakage events will be be handled automatically by the neural network decoder, based on the signature they leave in the detector data.

\section{Graph neural network decoder}
\label{sec:GNN_decoder}

A graph neural network (GNN) can be viewed as a trainable message passing algorithm, where information is passed between the nodes through the edges of the graph and processed through a neural network~\cite{kipf2016semi,wu2020comprehensive,dwivedi2020benchmarking}. 
The input is data in the form of a graph $G=(V,E)$, with a set of nodes $V=\{i\,|\, i=1,..,N\}$ and edges 
$E=\{(i,j) \,|\, i\neq j\in V\}$, which is annotated by $n$-dimensional node feature vectors $\vec{X}_i$ and edge weights $e_{ij}$. The data flow for our GNN-implementation is outlined in Figure \ref{fig:GNN_scheme}, with input in the form of an annotated detector graph and output in the form of two binary predictions.  
The basic building blocks are the message passing graph convolutional layers which take a graph as input and output an isomorphic graph with transformed feature vectors. 
Specifically, in this work we have used a standard graph convolution~\cite{morris2021weisfeiler} where for each node $i$ the  $d_{in}$-dimensional feature vector $\vec{X}_i$ is transformed to new feature vector $\vec{X}'_i$ with dimension $d_{out}$ according to 
\begin{equation}
    \vec{X}'_i=\sigma\left ( W_1\vec{X}_i+W_2\sum_j e_{ij} \vec{X}_j+\vec{b}\right )\,,
    \label{Eq:graphconv}
\end{equation}
where non-existent edges are indicated by $e_{ij}=0$. Here $W_1$ and $W_2$ are $d_{out}\times d_{in}$ dimensional trainable weight matrices, $\vec{b}$ is a $d_{out}$-dimensional trainable bias vector. The non-linear activation function, $\sigma$, acts element-wise, outputting the new feature vector. A standard form, used in this work, is the rectified linear unit, $\sigma(x)=\text{ReLU}(x)=\text{max}(0,x)$.  

For the task at hand, which is graph classification, a number of subsequent graph convolutions are followed by a pooling layer that contracts the information to a single vector, a graph embedding, which is independent of the dimension of the graph. 
In this work, we use a simple mean-pooling layer 
\begin{equation}
    \vec{X}_{\text{mean}}=N^{-1}\sum_{i}\vec{X}_i\,,
\end{equation}
where $N$ is the number of nodes in the graph. 
For the classification we use two structurally identical, but independent, multi-layer perceptrons (MLP), i.e.\ standard dense feed-forward neural networks, where each layer acts acts as  
\begin{equation}
    \vec{X}'=\sigma(W\vec{X}+\vec{b})\,,
\end{equation}
with a trainable weight matrix $W$ and bias vector $\vec{b}$. The input to the two MLPs is  
the pooled output from the graph convolution layers, $\vec{X}_{\text{mean}}$. Each MLP ends with a single node with sigmoid activation, $\sigma(x)=1/(1+e^{-x})\in [0,1] $, that acts as a binary classifier. 
The weights and biases of the complete network are trained using stochastic gradient descent with a loss function which is a sum of the binary cross entropy loss of the network output with respect to the binary labels.  
Since the experimental data, or simulated data, only has one of the two binary labels ($\lambda_Z$, $\lambda_X$) for each complete detector graph, gradients are only calculated for the provided label. 

To avoid overfitting to the training data we employ two different approaches depending on the amount of available data. In using experimental data from \cite{google2023suppressing}, we use a two-way split into a training set and a test set. To avoid diminishing the training data further, we do not use a validation set, and instead train for a fixed number of epochs. We observe (see Figure~\ref{fig:training_rep}) that the test accuracy does not change significantly over a large number of epochs, even though the network continues to overfits.  

For the case with simulated  data (Figure \ref{fig:training_accuracy}), we avoid overfitting by not reusing data. Each batch of the training data consists of freshly generated, labeled detector graphs. A fixed test set is used to gauge the performance.

The GNN training and testing is implemented in PyTorch Geometric~\cite{fey2019fast}, simulated data is generated using Stim~\cite{gidney2021stim}, the MWPM decoding results use PyMatching~\cite{higgott2021pymatching,pymatchingv2}, and the belief-matching results uses the code provided with \cite{Higgott2022beliefmatching}. 
The Adam optimizer is used for stochastic gradient descent, using manual learning rate decrements when the training accuracy has leveled out. Details on the training procedure can be found in Appendix~\ref{sec:App_A}. Several other graph layers were experimented with, including graph attention for both convolutions~\cite{velivckovic2017graph} and pooling~\cite{lee2019selfattention,knyazev2019understanding}, as well as top$_k$ pooling~\cite{gao2019graph,cangea2018sparse}. 
These were found not to improve results. 
The width and depth of the final network was arrived at after several rounds of iterations, but no systematic ablation studies were done. 
We expect that larger code distances, i.e., larger graphs, will require scaling up the network, following the increased complexity of the decoding problem. We use a fixed-size network for $d\leq 7$, and a somewhat larger network for $d=9$  (See also Sec.~\ref{sec:time_complexity})  

\subsection{Data structure}
\label{sec:data_structure}
As discussed previously the data is in a form $D=(\{V_{Z}\},\{V_{X}\},\lambda_{Z/X})$, consisting of a set of detectors $V_{Z/X}$, specified by a space-time coordinate, together with a binary label. 
Based on this we construct a single graph. 
Each node corresponds to a detector event, and is annotated by a 5-vector (for the surface code with circuit-level noise) $\vec{X}=(b_1, b_2, x, y, t)$ containing the space-time coordinate $(x,y,t)$ and two exclusive binary (one-hot encoded) labels with $\vec{b}=(1,0)$ for an X-stabilizer and $\vec{b}=(0,1)$ for a Z-stabilizer. (The encoding of the type of stabilizer may be superfluous, as it can be deduced from the coordinate.)
We initially consider a complete graph, with edge weights given by the euclidean distance between the detectors, $e_{ij}=1/\sqrt{(x_i - x_j)^2+(y_i - y_j)^2+(t_i - t_j)^2}$.  The edge weights give a rough measure of the likelihood that two detectors are triggered due to the same error or set of errors and are used to prune edges in the graph. Lower weight edges are removed, leaving only a fixed maximal node degree, reducing the size of each data point such that it grows linearly with the number of nodes. The pruning using euclidean distance is efficiently implemented in the integrated data generation and training pipeline as a data-preprocessing step. 

\section{Results}
\label{sec:results}
The GNN based decoder has been implemented, trained, and tested on the surface code and the repetition code. The main focus is on using simulated data or experimental data, presented in \ref{sec:stim_results} and \ref{sec:rep_results}, respectively. We also present some results on the surface code with perfect stabilizers,  \ref{sec:perfect_results}, where we are able to train the network for larger code distances. 

\subsection{Surface code with circuit-level noise }
\label{sec:stim_results}
We use Stim to generate data with circuit-level noise. Simulated circuits use standard settings for the surface code, containing Hadamard single qubit gates, controlled-Z (CZ) entangling gates, and measure and reset operations. All of these operations, and the idling, contain Pauli noise, scaled by an overall error rate $p$ (see Appendix \ref{sec:App_B}.) Datasets are generated in batches of varying sizes (see Appendix \ref{sec:App_A}), that each give a single gradient descent update of the neural network weights. For presentation purposes the batches are grouped into epochs containing $10^7$ data points in total, after which the test accuracy is evaluated. As discussed previously, to eliminate overfitting to the training data, no data is reused. This is feasible as the data generation is very fast.

\begin{figure}
\centering
\includegraphics[width=\linewidth]{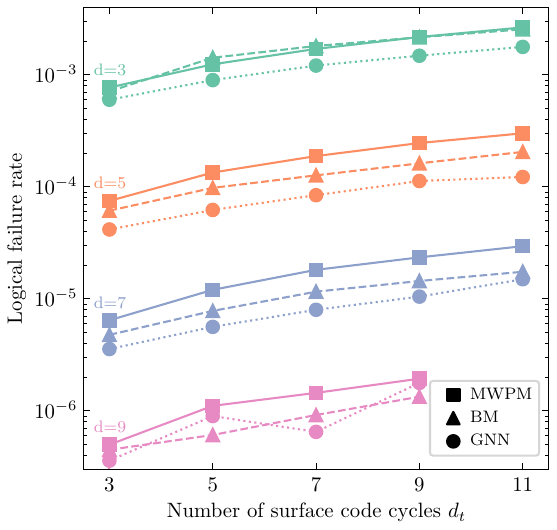}
\caption{Logical failure rate versus number of rounds of stabilizer measurements $d_t$, with simulated circuit-level noise~\cite{gidney2021stim} (error rate $p=1\cdot 10^{-3}$), on the surface code. Comparing Graph neural network (GNN) decoder to MWPM decoder~\cite{higgott2021pymatching} and belief-matching (BM) decoder~\cite{Higgott2022beliefmatching}. Each data point is evaluated over $10^8$ samples ($10^7$ for $d < 7$). Error bars are smaller than the markers.}
\label{fig:circuit_level}
\end{figure}

\begin{figure}
\centering
\includegraphics[width=\linewidth]{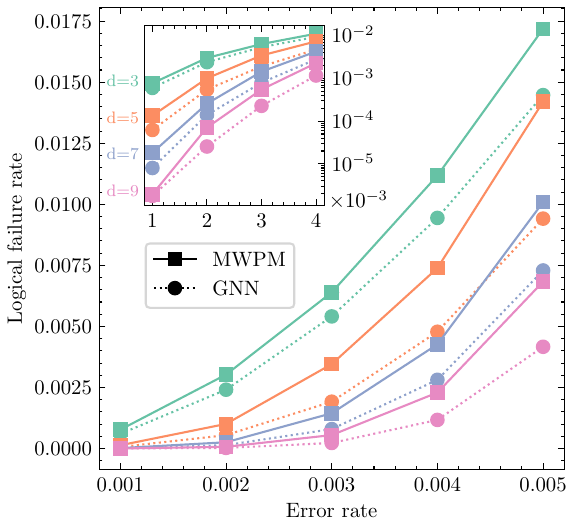}
\caption{Logical failure rate versus error rate $p$, with simulated circuit-level noise, on the surface code with code distance $d$ and $d_t=d$ stabilizer measurement cycles.  Else as in Fig.~\ref{fig:circuit_level}}
\label{fig:circuit_level_vs_p}
\end{figure}

Figure \ref{fig:circuit_level} shows test results evaluated at $p=1.0\cdot 10^{-3}$ for decoders trained with data using an even mix of error rates $p=\{1.0,2.0,3.0,4.0,5.0\}\cdot 10^{-3}$ and memory-Z experiments. The logical failure rate is thus approximately 50\% of the true failure rate (up to correlations between failures in $X_L$ and $Z_L$), but consistent with the type of data that would be experimentally accessible. (We have also tried training and testing with a mix of memory-Z and memory-X experiments, which works as well but takes longer to train to the same accuracy.) The rationale for using larger error rates during training is to include a relatively larger fraction of graphs with many nodes, under the assumption that these will generally be harder to decode. We compare to MWPM and belief-propagation augmented MWPM (belief-matching). Both these decoders use the information provided by the simulated error model to optimize edge weights on the decoding graph, where the BM algorithm additionally propagates information within and between the Z- and X-detector graphs for each instance. Despite the fact that the GNN decoder uses only the data provided by the simulated measurements, we find that with sufficient training the GNN decoder outperforms the matching decoders. For the largest code-distance considered, $d=9$, a larger network was used (see Appendix \ref{sec:App_A}), and the training has not converged to consistently outperform BM for all cycle depths considered. Figure \ref{fig:circuit_level_vs_p} also shows the performance of the GNN under varying error rates versus MWPM. We find that the networks have good performance within the whole range of error rates over which it was trained. 

\begin{figure}
\centering
\includegraphics[width=\linewidth]{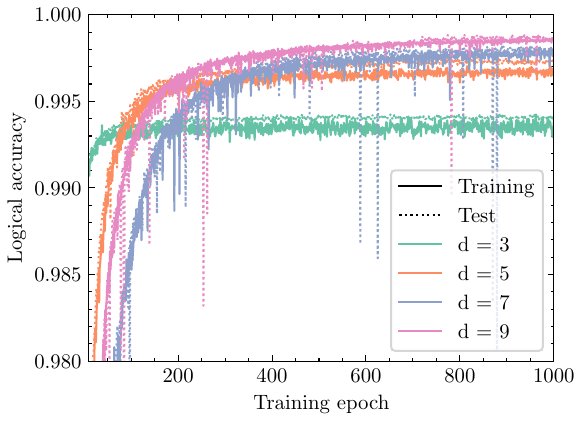}
\caption{GNN training and test accuracy versus number of training epochs for circuit-level noise, comparing different code distances ($d_t = d$ ). Each epoch amounts to training on a freshly generated dataset with $10^7$ samples using an error rate randomly selected from $p = [0.001, 0.002, ..., 0.005]$. The test set is a fixed dataset of the same type containing $5\cdot 10^4$ data points. The discrepancy with  accuracies in Figure \ref{fig:circuit_level} is due to the exclusion of empty graphs (trivial syndromes) from the training data.}
\label{fig:training_accuracy}
\end{figure}

A different network is trained for each code distance $d$ and for each number of rounds of stabilizer measurements $d_t$. Figure \ref{fig:training_accuracy} shows a representative plot of the training and test accuracy, evaluated on the mixed error rate dataset. No data is reused, which implies that the network cannot overfit and that the test accuracy closely follows the training accuracy. Further details are given in Appendix \ref{sec:App_A}.


\subsection{Repetition code using experimental data}
\label{sec:rep_results}
Having trained GNN based decoders on simulated data in the previous section, we now turn to real experimental data. We use the public data provided together with \cite{google2023suppressing}. This contains data on both the $d=3$ and $d=5$ surface codes as well as the $d=25$ bit-error correcting repetition code. All datasets are of the form described in \ref{sec:memory_exp}, thus readily transferred to the annotated and labeled graphs used to train the GNN, as described in \ref{sec:data_structure}. The datasets contain approximately $10^6$ data points for the different codes, code distances, and varying number of stabilizer rounds. 

Our attempts to train a GNN on the data provided for the various implementations of surface code were generally unsuccessful. While it gave good results on the training data, the logical failure rate on the test set was poor. Given the fact that on the order of $10^9$ data points were used for the simulated circuit-level noise on the surface code (\ref{sec:stim_results}), it is not surprising that the significantly smaller dataset turned out to be insufficient. The network cannot achieve high accuracy without overfitting to the training data given the relatively small dataset. 

\begin{figure}
\centering
\includegraphics[width=\linewidth]{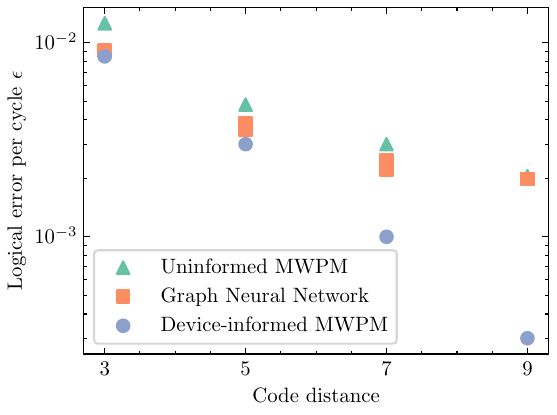}
\caption{Decoding experimental data \cite{google2023suppressing} on the repetition code with code distance $d$, over 50 rounds of stabilizer measurements. Comparing GNN decoder, using a dataset containing $(26-d)\cdot 5\cdot 10^7$ graphs, with a  MWPM decoder with ``device-optimized'' edge weights (\cite{google2023suppressing}) and a simple model-free MWPM decoder with 1-norm edge weights. The training-test split of the dataset is 99 to 1, and the logical failure rate is mapped to an error rate per round. Results for two different random training-test splits are shown.}
\label{fig:repetiotion_code_results}
\end{figure}

\begin{figure}
\centering
\includegraphics[width=\linewidth]{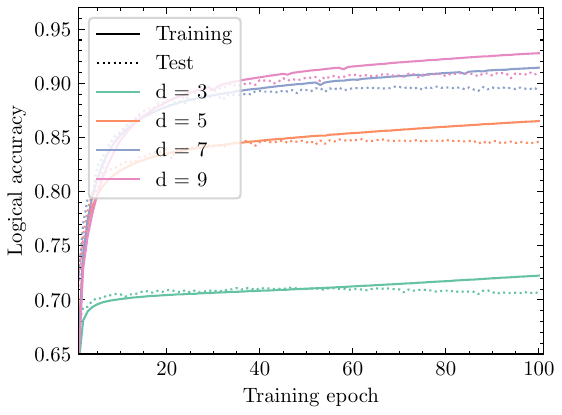}
\caption{Training curves for the GNN decoder on repetition code, following Figure \ref{fig:repetiotion_code_results}.  Each epoch trains through the whole training set, which eventually leads to overfitting, where the training accuracy starts to significantly surpass the test accuracy. To maximize the amount of training data, no validation set was used. No early stopping was implemented in order to avoid optimizing results to the test set.}
\label{fig:training_rep}
\end{figure}

For the repetition code, the data which is provided is of a single type, for a $d=25$ code measured over $d_t=50$ rounds.  Each round thus contains the measurement of 24 ancilla qubits for the $ZZ$ stabilizers of the two neighboring data qubits along a one-dimensional path. As done in \cite{google2023suppressing} this data can be split up into data for smaller repetition codes, by simply restricting to stabilizers over a subset of $d$ subsequent data qubits. In this way the dataset can be increased by a factor $25-(d-1)$, and used to train a single GNN for each code distance. It should be noted that this is suboptimal, compared to generating the same amount of data on single distance $d$ device, as variations in the performance of the constituent qubits and gates will be averaged out in the dataset. 
Nevertheless, using this scheme we successfully trained GNN decoders for short distance repetition codes, with test accuracies shown in Figure \ref{fig:repetiotion_code_results}. Results for (what we refer to as) ``Device-optimized MWPM'' is taken from \cite{google2023suppressing}. The GNN decoder performs almost on par with this sophisticated matching decoder for $d=3$. As expected, the relative performance deteriorates with increased code distance. We expect that we would need more training data for larger code distance, but instead we have access to less.    

As the comparison with the matching decoder that uses a device specific error model may be biased compared to using training data from different devices, as mentioned above, we also give results for an ``{u}niformed'' matching decoder with edge weights based on the 1-norm distance between space-time coordinates. It may also be noted that using MWPM corresponds to a near optimal decoder for the repetition code, at least for the case of phenomenological measurement noise where it is equivalent to bit-flip error surface code. 
This is in contrast to the surface code, for which MWPM is suboptimal, even in the case of perfect stabilizers. Thus, outperforming MWPM for the repetition code may be more challenging than for the surface code. 

\subsection{Surface code with perfect stabilizers}
\label{sec:perfect_results}
To complement the results on circuit-level noise and experimental data we have also trained the GNN decoder on the surface code with perfect stabilizers under depolarizing noise. The same network (see Appendix \ref{sec:App_A}) is used as for circuit-level noise, but trained at $p=[0.01, 0.05, 0.1, 0.15]$. For this problem, both labels, corresponding to logical bit- and/or phase-flips, are used for training and testing.

\begin{figure}
\centering
\includegraphics[width=\linewidth]{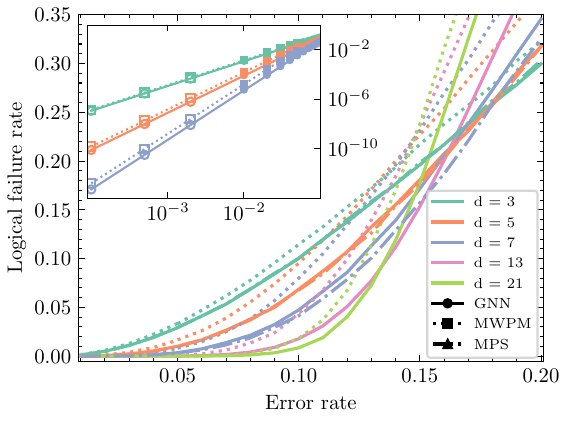}
\caption{Decoding the rotated surface code with perfect stabilizers and code distance $d$. Logical failure rate versus error rate $p$, for depolarizing noise, evaluated over failures with respect to both $X_L(\lambda_X)$ and $Z_L(\lambda_Z)$. Comparing the GNN decoder with MWPM decoder that has full information of the data-generating error model. Each data point is evaluated over $10^5$ data points ($10^8$ for $p<10^{-2}$). Dashed lines is the accuracy using a matrix product state (MPS) decoder~\cite{qecsim} at code distances 3 to 7. Inset shows low-$p$ failure rates for $d\leq 7$, where open markers are based on sampling only the lowest weight errors that fail.}
\label{fig:perfect_stabilizer}
\end{figure}

Results up to code distance $d=21$ are shown in Figure \ref{fig:perfect_stabilizer} and found to significantly outperform MWPM. We also compare to a tensor network based~\cite{qecsim} maximum likelihood decoder (MLD), showing that for code distance $d\leq 5$ the GNN decoder has converged to the level of being an approximate MLD. For very low error rates and larger code-distances $d>7$, we find that the networks still fail for some low weight errors that should be correctable, making the asymptotic behavior worse than MWPM. Nevertheless, we expect that more training, and training tailored to low error rates, could resolve this. 

We do not attempt to derive any threshold for the GNN decoder. Given a sufficiently expressive network we expect that the decoder would eventually converge to a maximum likelihood decoder, but in practice the accuracy is limited by the training time. It gets progressively more difficult to converge the training for larger code distances, which means that any threshold estimate will be a function of the training time versus code distance. In fact, in principle, since the threshold is a 
$d\rightarrow\infty$ quantity, we would not expect that a supervised learning algorithm can give a proper threshold if trained separately for each code distance, as is done in this work. Using GNN's (as opposed to a network acting on a fixed size grid) it is in principle quite natural to use the same network to decode any distance code, as the data objects (detector graphs) have the same structure. We have investigated training the same network for different code distances and different number of rounds. This shows some promise, but has not achieve accuracy levels that can match MWPM.    

\subsection{Scalability}
\label{sec:time_complexity}
We are limited to relatively small codes in this work. For the repetition code using experimental data, it is quite clear that main limitation to scaling up the code distance is the size of the available dataset. For the surface code using simulated data it is challenging to increase the code distance while still surpassing MWPM. As the logical failure rates decrease exponentially with code distance, the test accuracy of the supervised training needs to follow. One way to counter this is to increase the number of stabilizer cycles, $d_t$, but this also increases the graph size, making the training more challenging from the perspective of increased memory requirements as well as the increased complexity of the data. 

Nevertheless, it is interesting to explore the intrinsic scalability of the algorithm, by quantifying how the decoding time using a fixed size GNN scales with the code size. Here we present results on the decoding time per syndrome for the surface code, as a function of code volume $d^2d_t$, at fixed error rate, compared to PyMatching 2~\cite{pymatchingv2}. The network is fixed to the smaller network  described in Appendix \ref{sec:App_A}. In line with expectations, the GNN inference scales approximately linearly with the code volume, i.e.\ average graph size, $T\sim d^2d_t$. The number of matrix operations per graph convolutional layer, following Equation \ref{Eq:graphconv}, is proportional to the number of nodes times the number of edges. The number of layers is fixed, multiplying this by a constant factor. The feature vector pooling is proportional to the number of nodes, whereas the subsequent dense network classifiers are independent of the graph size. We find that inference scales slightly better than the highly optimized matching decoder. However, several caveats are in order. 1) The size of the GNN is fixed. Larger code will require scaled up networks, unless the error rate is scaled down accordingly  2) The network has not been trained on code distances larger than $d=9$. It is only a test of the decoding time, not the accuracy.  3) Data (for both algorithms) is batched for fast inference. Treating batched data doesn't seem viable for real time decoding. Moving to other types of hardware (see, e.g.\ \cite{Barber_2025}), such as field-programmable gate arrays (FPGA) or an application-specific integrated circuits (ASIC), will be necessary for real-time decoding of superconducting devices, requiring $\mu$s per cycle decoding times, using neural networks. 

\section{Conclusion and Outlook}
\label{sec:Conclusion}

In this paper we develop a model-free, data-driven, approach to decoding quantum error correcting stabilizer codes, using graph neural networks. 
A real or simulated memory experiment is represented as a single detector graph, with annotated nodes corresponding to the type of stabilizer and its space-time coordinate, and a binary label corresponding to whether or not a logic bit-flip has occurred. The maximal node degree is capped by cropping edges between distant nodes.   
The data is used to train a convolutional GNN for graph classification, with classes corresponding to logical Pauli operations, and used for decoding. 
We show that we can use experimental and simulated data, for the repetition code and surface code respectively, to train a decoder with logical failure rates on par with minimum weight perfect matching, despite the latter having detailed information about the underlying real or simulated error channels. 
The use of a graph structure provides an efficient way to store and process the syndrome data. Training the GNN requires significant amounts of training data, but as shown in the case of simulations, data can be produced in parallel with training. 
Network inference, i.e., using the network as a decoder, is fast, scaling approximately linearly with the space-time dimension of the code.     


As an extension of this work there are several interesting possibilities to explore. One example is to use a GNN for edge weight generation within a hybrid algorithm with a matching decoder (similarly to \cite{Chen2021ExponentialCorrection}). This would depart from the pure data-driven approach pursued in this paper, with peak performance limited by the matching decoder, but with the potential advantage of requiring less data to train. An alternative to this, to potentially improve performance and lower data requirements, is to use device specific input into edge weights, or encode soft information on measurement fidelities into edge or node features.     

Going beyond the most standard codes considered in this paper, we expect that any error correcting code for which labeled detector data can be generated can also be decoded with a GNN. 
This includes Clifford-deformed stabilizer codes \cite{Tuckett2018, Ataides2021XZZX, Dua2022, Tiurev2022, Huang2022}, color codes \cite{Bombin2006colorcode, Bombn2015}, hexagonal stabilizer codes \cite{Wootton2015,Wootton2021,Srivastava2022xyzhexagonal, Hetenyi2023} and quantum low-density parity check (LDPC) codes~\cite{PRXQuantum.2.040101,panteleev2022asymptoticallygoodquantumlocally,Bravyi_2024}, where syndrome defects are not created in pairs, but potentially also Floquet type codes~\cite{Haah2021BoundariesDGLT,Kesselring2022FloquetColorCode}.
In addition, heterogeneous and correlated noise models~\cite{iOlius_2022,tiurev2023correcting} would also be interesting to explore, where in particular the latter is difficult to handle with most standard decoders. Adding soft information, e.g.\ the estimated reliability of a stabilizer measurement, is also a natural next step for this type of decoder~\cite{pattison2021improvedquantumerrorcorrection}. The software code for the project can be found at \cite{git_GNN_decoder}.

Shortly after posting the preprint for this paper, related work, \citet{varbanov2023neural}, using a recurrent neural network architecture was presented. That network was trained on simulated data and tested on  experimental data \cite{google2023suppressing}, but did not implement training exclusively on experimental data. Related work has subsequently been presented in \citet{bausch2024learning}, using a transformer augmented recurrent architecture.   


\begin{acknowledgments}

We acknowledge financial support from the Knut and Alice Wallenberg Foundation through the Wallenberg Centre for Quantum Technology (WACQT), and the Marianne and Marcus Wallenberg foundation through EDU-WACQT. Computations were enabled by resources provided by the National Academic Infrastructure for Supercomputing in Sweden (NAISS) and the Swedish National Infrastructure for Computing (SNIC) at Chalmers Centre for Computational Science and Engineering (C3SE), partially funded by the Swedish Research Council through grant agreements no. 2022-06725 and no. 2018-05973, as well as by the Berzelius resource provided by the Knut and Alice Wallenberg Foundation at the National Supercomputer Centre. We thank Viktor Rehnberg and Hampus Linander for input on the neural network training. This work was also supported by the Dutch National Growth Fund (NGF), as part of the Quantum Delta NL programme.

\end{acknowledgments}

\appendix
\section{GNN architecture and training}
\label{sec:App_A}
\begin{figure}
\centering
\includegraphics[width=\linewidth]{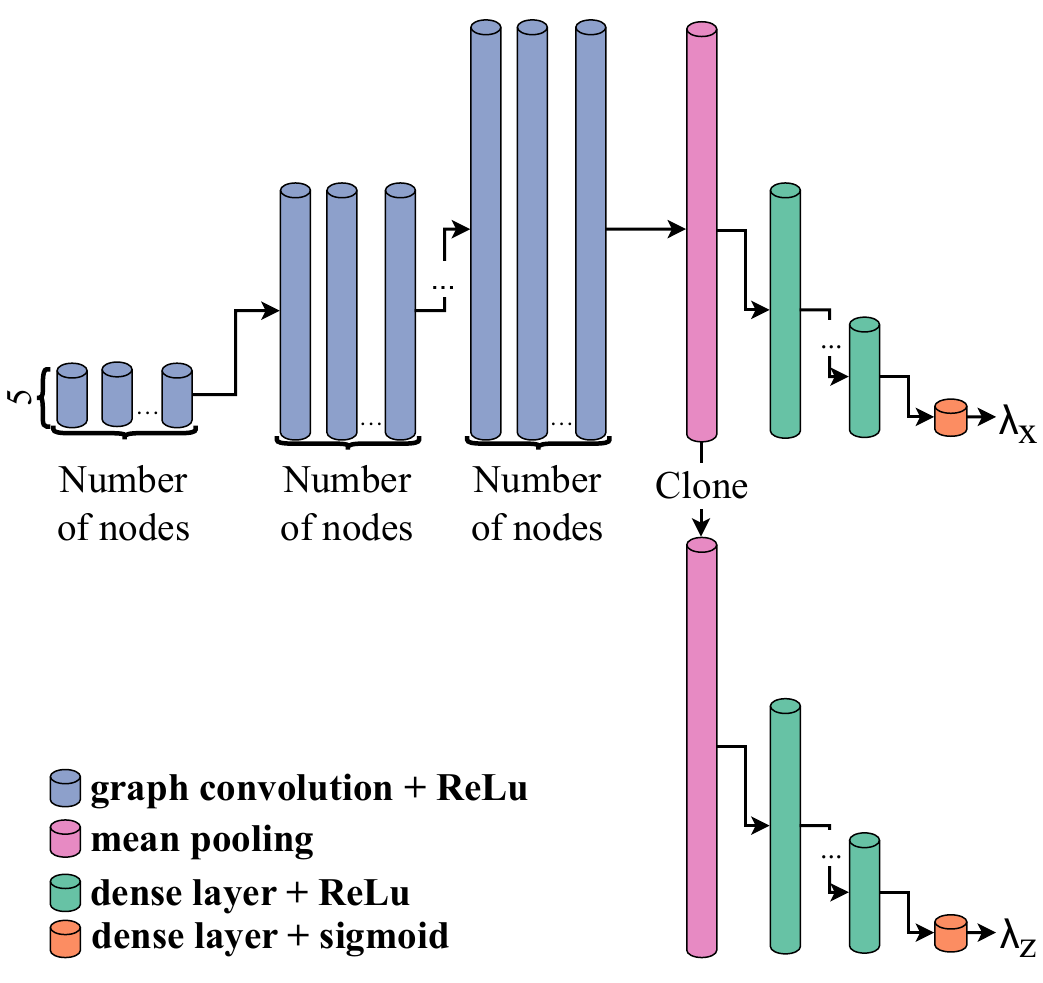}
\caption{Schematic of the GNN  architecture, with details in Table \ref{fig:layer_dimensions}. The same architecture is used for all the results, except that for the repetition code there is only one output head. The input dimension is two (2D space-time coordinate) for the repetition code and four (two types of stabilizers, and 2D spatial coordinate) for the surface code with perfect stabilizers.      
}
\label{fig:network_architecture}
\end{figure}

Figure \ref{fig:network_architecture} displays the architecture of the GNN decoder. The node features are sent through 7 subsequent graph convolutional layers (Equation \ref{Eq:graphconv}). The node features are passed through a rectified linear unit (ReLU) activation function (which corresponds to chopping negative values) after each layer. 
After the graph convolutional layers, the node features from all nodes are pooled into one high-dimensional vector by computing the mean across all nodes. This vector is then cloned and sent to two identical fully connected neural networks. Both heads consist of several dense layers which map the pooled node feature vector down to one real-valued number which is mapped to the range 0 to 1 through a sigmoid function. The input and output dimension $d_{in}$ and $d_{out}$ of the graph convolutional and dense layers can be found in Table \ref{fig:layer_dimensions}. 

\begin{table}[h]
    \centering
    \begin{minipage}{0.25\linewidth}
        \centering
        \begin{tabular}{|c|c|c|}
            \toprule
            Layer        & $d_{in}$ & $d_{out}$ \\
            \midrule
            GraphConv\textsubscript{1} & 5   & 32  \\
            GraphConv\textsubscript{2} & 32  & 128 \\
            GraphConv\textsubscript{3} & 128 & 256 \\
            GraphConv\textsubscript{4} & 256 & 512 \\
            GraphConv\textsubscript{5} & 512 & 512 \\
            GraphConv\textsubscript{6} & 512 & 256 \\
            GraphConv\textsubscript{7} & 256 & 256 \\
            Dense\textsubscript{1}   & 256 & 256 \\
            Dense\textsubscript{2}   & 256 & 128 \\
            Dense\textsubscript{3}   & 128 & 64  \\
            Dense\textsubscript{4}   & 64  & 1   \\
            \bottomrule
        \end{tabular}
    \end{minipage}
    \hfill
    \begin{minipage}{0.45\linewidth}
        \centering
        \begin{tabular}{|c|c|c|}
            \toprule
            Layer        & $d_{in}$ & $d_{out}$ \\
            \midrule
            GraphConv\textsubscript{1} & 5   & 32  \\
            GraphConv\textsubscript{2} & 32  & 128 \\
            GraphConv\textsubscript{3} & 128 & 256 \\
            GraphConv\textsubscript{4} & 256 & 512 \\
            GraphConv\textsubscript{5} & 512 & 512 \\
            GraphConv\textsubscript{6} & 512 & 512 \\
            GraphConv\textsubscript{7} & 512 & 512 \\
            Dense\textsubscript{1}   & 512 & 512 \\
            Dense\textsubscript{1}   & 512 & 256 \\
            Dense\textsubscript{3}   & 256 & 128 \\
            Dense\textsubscript{4}   & 128 & 64  \\
            Dense\textsubscript{5}   & 64  & 32   \\
            Dense\textsubscript{6}   & 32  & 16   \\
            Dense\textsubscript{7}   & 16  & 1   \\
            \bottomrule
        \end{tabular}
    \end{minipage}
    \caption{Overview of the input and output dimensions of the graph convolutional and dense layers of the GNN decoder. Left: $d\leq 7$, total number of parameters: $1.36\cdot 10^6$. Right: $d=9$, total number of parameters: $2.35\cdot 10^6$.}
    \label{fig:layer_dimensions}
\end{table}

Networks are trained on NVIDIA Tesla A100 HGX GPU's using the pytorch geometric knn module to generate graphs in parallel. For gradient descent, samples are batched in batches of sizes ranging from $6 \cdot 10^3$ for the biggest code instances ($d = d_t = 9)$ to $26 \cdot 10^3$  for the smallest code instances ($d = d_t = 3)$. The batch sizes are chosen to fully utilize the GPU during training. The learning rate is set to $10^{-4}$ and decreased manually to $10^{-5}$, whenever the validation accuracy reached a plateau. The training scripts, trained models and details on all hyper-parameters are available at \cite{git_GNN_decoder}.

\section{Stabilizer circuits and error model for circuit-level noise}
\label{sec:App_B}
\begin{figure}
\centering
\includegraphics[width=\linewidth]{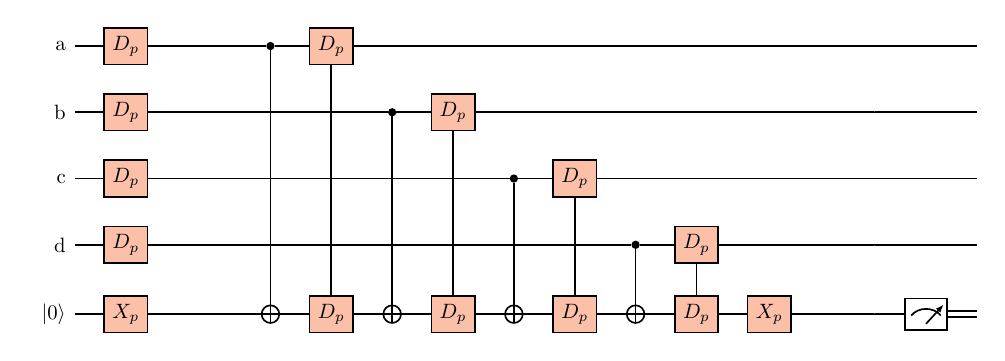}
\caption{Quantum circuit for measuring the weight-four stabilizer $Z_\text{abcd}$ under circuit-level noise.}
\label{fig:z_stabilizer_circuit}
\end{figure}

\begin{figure}
\centering
\includegraphics[width=\linewidth]{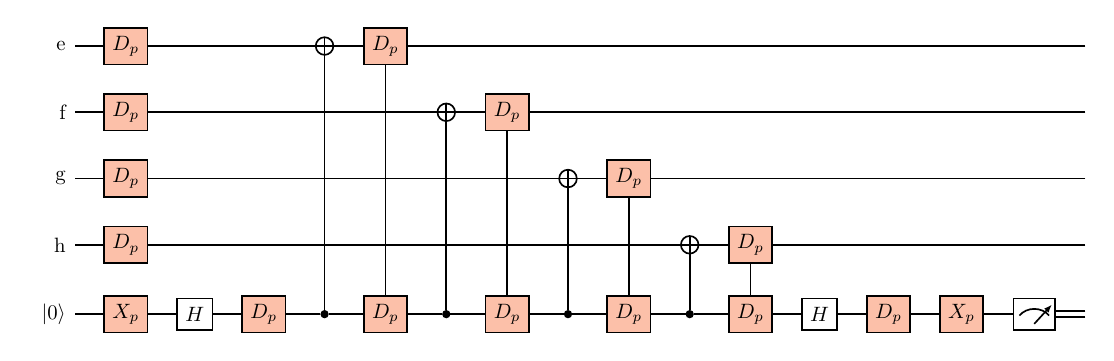}
\caption{Quantum circuit for measuring the weight-four stabilizer $X_\text{efgh}$ under circuit-level noise.}
\label{fig:x_stabilizer_circuit}
\end{figure}

Quantum circuits for weight-four $Z$- ($X$-) stabilizers of the surface code are displayed in Figure \ref{fig:z_stabilizer_circuit} (\ref{fig:x_stabilizer_circuit}). The gate set used for the stabilizer measurements consists of the Hadamard gate $(H)$, and the $CNOT$ gate. Under circuit-level noise, single-qubit depolarizing noise gate $D_p$ (which applies gate $\sigma_i, i \in \{X, Y, Z\}$ where any of the gates is applied with probability $p/3$, and $I$ with probability $1-p$) acts on the data qubits before each stabilizer measurement cycle and on each target qubit after single-qubit gates. Two-qubit depolarizing noise gates (which apply gate $\sigma_i \sigma_j, i,j \in \{I, X, Y, Z\}$, where $II$ is acted on with probability $1-p$, and the rest with probability $p/15$) act on the two qubits involved after every two-qubit gate.  Furthermore, each qubit suffers from reset- and measurement-error with probability $p$, displayed by operators $X_p$ when measuring and resetting in the computational basis. 

\newpage

\bibliography{ref}


\end{document}